\documentclass[10pt, conference, letterpaper]{IEEEtran}

\IEEEoverridecommandlockouts
\usepackage{cite}
\usepackage{amsmath,amssymb,amsfonts}
\usepackage{algorithmic}
\usepackage{graphicx}
\usepackage{textcomp}
\usepackage{xcolor}
\usepackage{caption}
\usepackage{subcaption}
\usepackage{multirow}
\usepackage{rotating}
\usepackage{tabularx}
\usepackage{booktabs}
\usepackage{afterpage}

\begin{document}

\title{An Experimental Study of Congestion Control in LEO Satellite Networks}
\author{
\IEEEauthorblockN{Mihai Mazilu$^{*}$, Aiden Valentine$^{*}$, Ian Wakeman$^{*\dagger}$ and George Parisis$^{*}$}
\IEEEauthorblockA{School of Engineering and Informatics, University of Sussex$^{*}$\\ Zhejiang Gongshang University$^{\dagger}$\\
\{m.mazilu, a.valentine, ianw, g.parisis\}@sussex.ac.uk}
}
\maketitle

\begin{abstract}

Low Earth Orbit (LEO) satellite networks introduce unique congestion control (CC) challenges due to frequent handovers, rapidly changing round-trip times (RTTs), and non-congestive loss. Developing effective transport for these environments requires a clear, comparable understanding of the performance profiles of both state-of-the-art Internet CC and emerging LEO-tailored designs, yet producing such profiles is challenging because LEO dynamics make experiments difficult to control, reproduce, and interpret. To address this, we conduct an emulation-driven evaluation of CC schemes in LEO networks, combining realistic orbital dynamics via the LeoEM framework with targeted Mininet micro-benchmarks that isolate specific conditions. We evaluate representative algorithms from three classes: loss-based Cubic, model-based BBRv3, and learning-based approaches Vivace and Astraea, alongside LEO-tailored SaTCP and LeoCC, across diverse scenarios. Results show that (1) handover-aware schemes can reclaim bandwidth but often at the cost of increased latency; (2) BBRv3 sustains high throughput with modest delay penalties, yet adapts slowly to abrupt RTT shifts; (3) learning-based schemes underperform under dynamic conditions, despite strong resistance to non-congestive loss; and (4) fairness degrades markedly under RTT asymmetry and multiple bottlenecks, particularly for human-designed CC schemes. These findings highlight limitations of today’s CC approaches and help inform the design of LEO-specific schemes.

\end{abstract}

\section{Introduction}
\label{intro}
Low Earth Orbit (LEO) satellite networks consist of thousands of interconnected satellites that form a global mesh orbiting around Earth. With an expected total bandwidth reaching hundreds of terabits per second, their capacity will rival the cumulative throughput of modern fibre-optic infrastructure \cite{21st_century_space_race}. Free-space laser links enable high-speed inter-satellite communication, operating in a vacuum at speeds approximately 47\% faster than in optical fibre cables \cite{delay_not_an_option}. As a result, LEO satellite networks can support lower latency communications than in terrestrial networks for distances greater than about 3000 km \cite{delay_not_an_option}. Deploying the IP stack in these networks requires careful consideration when sizing network buffers and deploying data transport protocols and congestion control (CC) algorithms so as not to increase latency through packet queues. For example, loss-based CC algorithms, such as Cubic \cite{cubic}, fill network buffers, particularly when operating in conjunction with large buffers \cite{hypatia,lucapaper_systematic_rlcc}. Even delay-based schemes, such as bottleneck bandwidth and round-trip time (BBR), operate with high buffer occupancy when multiple flows coexist \cite{hock_experimental_2017}. 

More broadly, LEO satellite networks present considerable challenges in the design of effective data transport and congestion control, given the dynamic nature of the topology,  routing paths \cite{delay_not_an_option}, and rain fade \cite{rain_fade}. These, in turn, result in (1) \textit{non-congestive latency variation and loss}, (2) \textit{transient hotspots} with associated latency inflation and packet loss, and (3) \textit{soft and hard handovers}. 

Existing LEO constellations commonly employ a \textit{+Grid} topology \cite{network_topology_design_at_27000kmh}, in which each satellite maintains four fixed laser inter-satellite links (ISLs): two to satellites in adjacent orbital planes and two to the nearest neighbours within the same plane. While these ISLs are fixed, the access links between satellites and ground stations or user terminals change continually as satellites move overhead \cite{hypatia}. This continual re-association triggers frequent handovers that can interrupt ongoing transmissions and induce packet loss \cite{satcp}. At the same time, satellite motion changes the end-to-end propagation distance, both between satellites and between satellites and the ground, leading to \textit{non-congestive} delay variation; when paths instead follow a bent-pipe (BP) structure, these latency swings can be amplified \cite{valentine_developing_2021,ma_network_2023}.  Routing decisions further interact with this dynamism: shortest-path routing over the LEO mesh can concentrate traffic onto a subset of links, creating \textit{transient hotspots} (especially along high-demand corridors shaped by uneven user distributions) that inflate queuing delay and increase loss \cite{21st_century_space_race,leoroutingchallanges}. Finally, rain fade can cause signal attenuation, resulting in slower data rates, increased latency, or temporary service interruptions \cite{rain_fade}. 

Various handover-aware transport solutions have been proposed to mitigate the disruption caused by frequent handovers. Broadly, these approaches aim to detect (or anticipate) handover events and adapt the sender’s behaviour to avoid excessive non-congestive loss. SaTCP \cite{satcp} and StarQUIC \cite{starquic} freeze the congestion window around handover periods, whereas the latest development, LeoCC \cite{leocc}, takes a more dynamic approach by explicitly accounting for LEO-induced path reconfigurations and resetting stale bottleneck estimates so the sender can quickly re-tune its rate to the new delay/capacity conditions. Although effective for the specific handover scenarios they target, these designs often come with drawbacks that have not yet been explored.

We argue that a clear understanding of how different classes of CC algorithms behave across the multiple challenges of LEO networks is necessary to achieve high performance and efficient use of network resources. Previous work has identified several pitfalls in evaluating CC algorithms \cite{lucapaper_systematic_rlcc, li2007experimental} such as not exercising CC, focussing on a narrow set of performance metrics, and evaluating CC in a limited range of network parameters. To date, there have been limited studies of CC in LEO satellite networks. Experiments with NewReno, Vegas, Cubic, and BBRv1 have been carried out on simulated LEO satellite paths \cite{hypatia, hypatia_experiments}.  Similarly, BBRv1, BBRv3, Cubic, and SaTCP \cite{satcp} have been studied with emulated single-path, single-flow scenarios \cite{satcp, starquic}, focussing primarily on handover-aware enhancements to TCP and QUIC. A broader spectrum of approaches have been explored in \cite{leocc}. 

In this paper, we advance the study of how existing congestion-control algorithms perform in LEO satellite networks. We evaluate loss-based schemes (TCP Cubic \cite{cubic} and SaTCP \cite{satcp}), model-based schemes (BBRv3 \cite{bbrv3} and LeoCC \cite{leocc}), and contemporary learning-based approaches, including Vivace \cite{vivace_uspace} and the reinforcement learning (RL) based Astraea \cite{astraea}. RL-based CC has been seen as an alternative to traditional human-derived schemes, promising adaptability under dynamic network conditions. It is therefore of great interest to understand how such CC schemes would behave if deployed on LEO satellite networks. Our study is based on network emulations. We use the LeoEM framework \cite{satcp} and Mininet-based micro-benchmarks \cite{mininet} to investigate key performance metrics, such as fairness, efficiency, and responsiveness. LeoEM \cite{satcp} enables us to accurately emulate delay variation and handovers in pre-specified LEO satellite network paths that use ISLs, or a bent-pipe (BP) topology. Then, we isolate and examine specific aspects of CC performance in a controlled and systematic fashion using micro-benchmarks. This allows for observing nuanced performance patterns that might be obscured in a more complex evaluation framework, such as LeoEM.

Key findings of our study include: (1) handover-aware loss-based schemes can reclaim bandwidth but at the cost of increased latency; (2) BBRv3 sustains high throughput with modest delay penalties, yet reacts slowly to abrupt RTT changes; (3) RL-based schemes underperform under dynamic conditions, despite being notably resistant to non‑congestive loss, though Astraea shows improved fairness due to its reward structure; (4) fairness degrades significantly with RTT asymmetry and multiple bottlenecks, especially in human-designed CC schemes.
\section{Related Work}
\label{related_work}

\noindent\textbf{Experimentation Frameworks}. 
The two principal simulation frameworks \cite{valentine_developing_2021,hypatia} use packet‑level simulators capable of supporting any networking stack permitted by their underlying models, providing both flexible network configurations and fully customizable satellite constellations.
More recently, several LEO emulation frameworks have been introduced. StarryNet \cite{starrynet} employs Docker containers to represent network components and relies on Docker’s bridge interface for link creation. Its scalability, however, is constrained to at most 1023 containers per machine, and the overhead of its Python-based design necessitates multiple high‑performance servers for larger experiments \cite{xeoverse}. Xeoverse \cite{xeoverse} reduces emulation cost by instantiating only the active links rather than all nodes, but despite the promise of this approach, the framework is not publicly available.

\noindent\textbf{CC in LEO Satellite Networks}.
In \cite{hypatia}, the authors assessed NewReno and Vegas over simulated LEO paths, identifying several shortcomings. Follow‑up work in \cite{hypatia_experiments} expanded this evaluation to include Cubic and BBRv1 on Starlink topologies, examining routing strategies and finding BBRv1 to deliver superior performance. The study in \cite{starquic} further emphasized the detrimental effects of lossy user-terminal handovers. By exploiting Starlink’s predictable reconfiguration patterns, the authors introduced a congestion‑window freeze mechanism for QUIC-based Cubic and BBRv3, analogous to SaTCP, and demonstrated its effectiveness.
In contrast to SaTCP and StarQUIC, LeoCC \cite{leocc} adopts a distinct strategy for detecting and responding to path reconfigurations. Its evaluation, based on Starlink measurements and trace‑driven emulation, shows enhanced resilience to LEO dynamics across a broad set of baselines. However, while such in‑the‑wild studies offer strong external validity, the limited controllability of operational networks makes it challenging to disentangle individual impairments—such as non‑congestive loss, re‑routing events, or multi‑bottleneck interactions—and to reproduce consistent scenarios across the wider set of congestion‑control families considered in this work.

Our work complements these efforts by offering a controlled, repeatable, emulation-driven evaluation that (i) systematically varies LEO-specific dynamics such as path reconfiguration, inter-RTT scenarios and multiple bottlenecks, and (ii) compares representative loss-based, model-based, and learning-based (including RL-based) CC designs under identical conditions, enabling clearer attribution of observed behaviours to underlying mechanisms.

In \cite{tcp_leo_starlink}, the authors conducted live measurements over Starlink using only CC schemes implemented within the Linux kernel, demonstrating that BBR (v1 and v3) achieves the highest throughput while the default Cubic under-utilises capacity. However, the paper largely focuses on aggregate performance outcomes like throughput and flow completion times and does not provide a detailed analysis of the underlying CC dynamics, such as how each of the tested congestion control algorithms react to variations in RTT, capacity, and queueing typical of LEO satellite links. In contrast, this work considers a smaller but more targeted subset of CC implementations chosen to be representative of newer developments in Internet CC and LEO-specific CC design. While this narrower scope reduces breadth, it enables a much deeper analysis of the selected CC approaches which can help inform future design decisions.  
\section{Methodology}
\label{methodology}

In \cite{lucapaper_systematic_rlcc}, the authors discuss trade-offs in experimentally evaluating CC using in-the-wild experiments, simulations, and emulations. Following their argument on balancing fidelity with flexibility and controllability of experiments, in this study, we adopted an emulation-driven approach using LeoEM and Mininet. Our setup leverages the Linux TCP/IP stack, which supports our selected traditional and RL-based CC schemes. Conducting in‑the‑wild experiments on LEO satellite networks is not practical and would hinder observability, control over network conditions, and reproducibility of results. Furthermore, existing LEO satellite network simulation frameworks \cite{hypatia}\cite{valentine_developing_2021} do not model handovers (as in \cite{satcp}) and the associated loss of connectivity and packet loss; they only model ground stations that perform lossless handovers without disrupting connectivity. 


\paragraph{\textbf{Selected CC approaches}} 
In this study, we experimented with Cubic, SaTCP, LeoCC, BBRv3, Vivace, and Astraea with the primary aim of covering a wide range of CC approaches, including traditional, loss-based, model-based, and RL-based ones. Hybla \cite{hybla} was designed for GEO satellite links, where RTTs are much larger than in Starlink-like LEO settings. We omit it because our experiments use much smaller RTTs than it was designed to address.

Cubic \cite{cubic} is the default congestion control in many modern operating systems and the most widely deployed CC scheme \cite{BRUHN2023109609}. SaTCP \cite{satcp} is a bespoke protocol for LEO satellite networks, and extends Cubic by introducing a freezing mechanism for its congestion window during handovers. SaTCP derives the freeze time by leveraging proactive handover reports from both user terminals and ground stations. BBR \cite{bbrv1} is a model-based algorithm that maximises throughput while minimising delay by continuously estimating the network’s bottleneck bandwidth and round-trip time and pacing its sending rate to match the estimated path capacity. Moreover, we included BBRv3 \cite{bbrv3}, BBR's latest iteration which introduces bug fixes related to convergence and parameter optimisation. 
LeoCC is a BBRv1-inspired congestion control algorithm that modifies BBRv1 to better suit LEO networks. In particular, it incorporates reconfiguration awareness, assisted by a companion application that notifies the sender when a reconfiguration occurs via the \textit{netlink} interface \cite{leocc}. Upon a reconfiguration, its model of the network will refresh, so it can discard stale bandwidth/RTT estimates and quickly re-tune its sending rate after LEO-induced path changes. 
Astraea \cite{astraea} uses multi-agent RL during training, improving upon previous work \cite{orca}, by explicitly integrating fairness, stability, and convergence into its reward function \cite{astraea}.

\paragraph{\textbf{Experimental setup}}
\label{par:experimentalSetup}
All experimentation is carried out using a BBRv3-enabled Linux kernel \cite{bbrthreekernel}, which we patched with the kernel component needed for Astraea. We employed LeoEM (Section \ref{sec:LeoEM}) to emulate real LEO satellite network paths. LeoEM is a lightweight, Mininet-based emulation framework that models user terminals, ground stations as well as hard and soft handovers. In our Mininet-based micro-benchmarks (Section \ref{sec:microbenchmarks}), we emulate different base RTT and non-congestive loss values, by configuring the Linux traffic control (\textit{tc}) queuing discipline \textit{netem}; bandwidth shaping is done using \textit{tbf} with drop-tail FIFO queues. This study was the result of $1650$ individual experiments. The source code for the frameworks and plotting as well as all data and individual experiments is available at \cite{data}. We have set the socket buffers (both send and receive) to a high value to prevent bottlenecks. 

\begin{figure*}[t]
    \centering
    \begin{subfigure}[t]{0.26\textwidth}
        \centering
        \includegraphics[width=\linewidth]{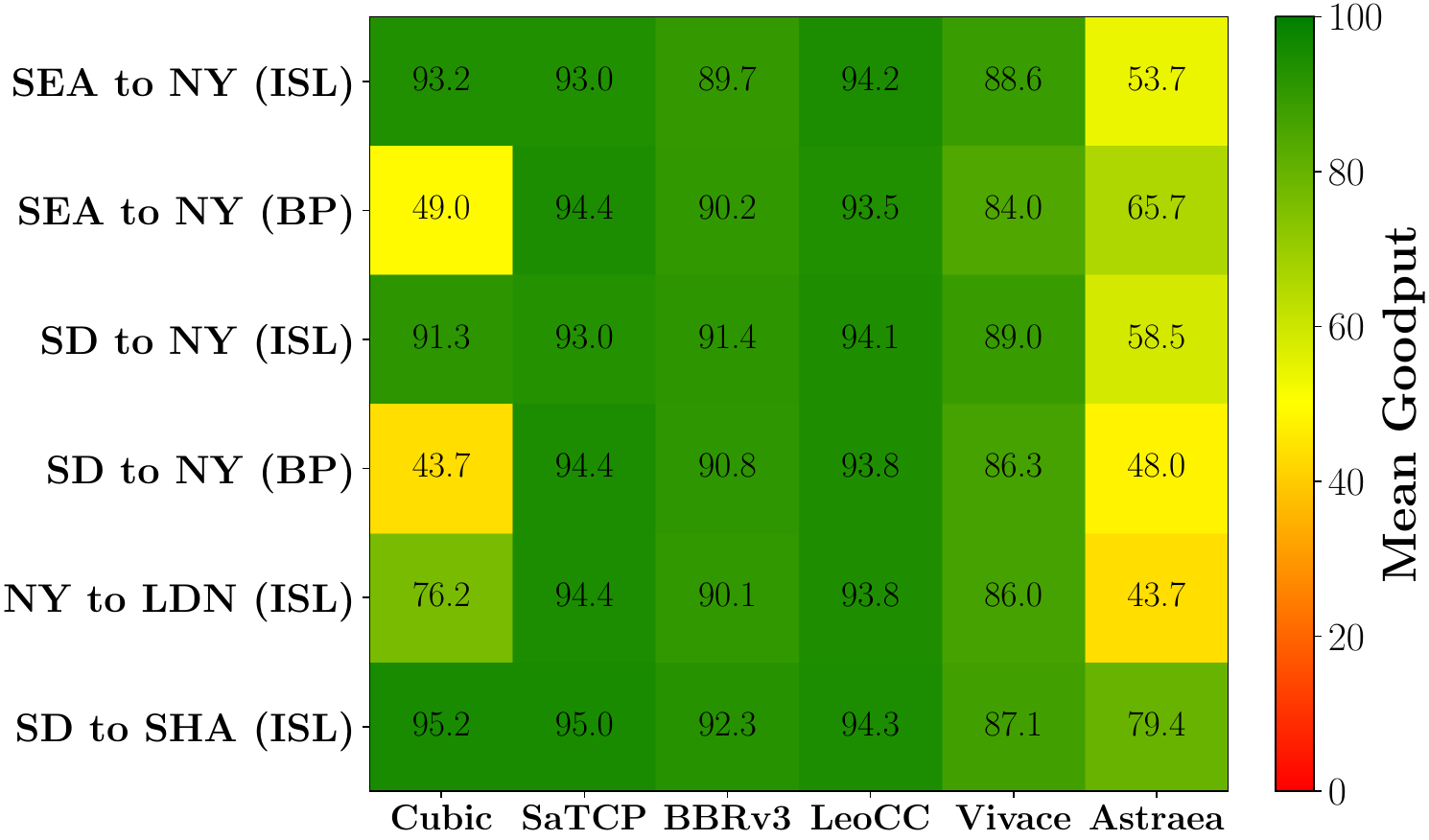}
        \vspace{-0.5cm}
        \caption{Normalised goodput}
        \label{fig:leoem_heatmap_goodput}
    \end{subfigure}
    \begin{subfigure}[t]{0.23\textwidth}
        \centering
        \includegraphics[width=\linewidth]
            {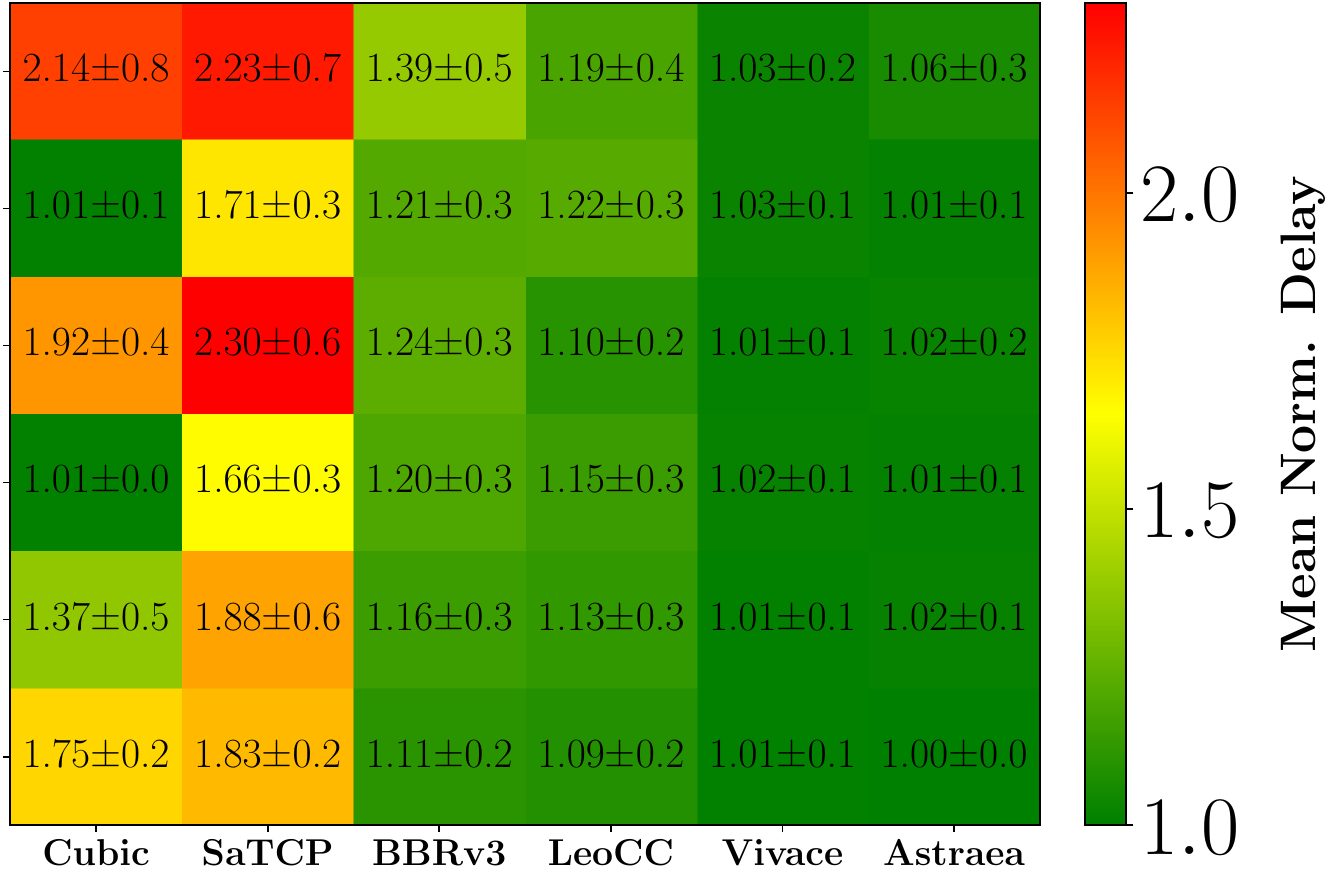}
        \vspace{-0.5cm}
        \caption{Normalised delay}
        \label{fig:leoem_heatmap_delay}
    \end{subfigure}
    \begin{subfigure}[t]{0.23\textwidth}
        \centering
        \includegraphics[width=\linewidth]
            {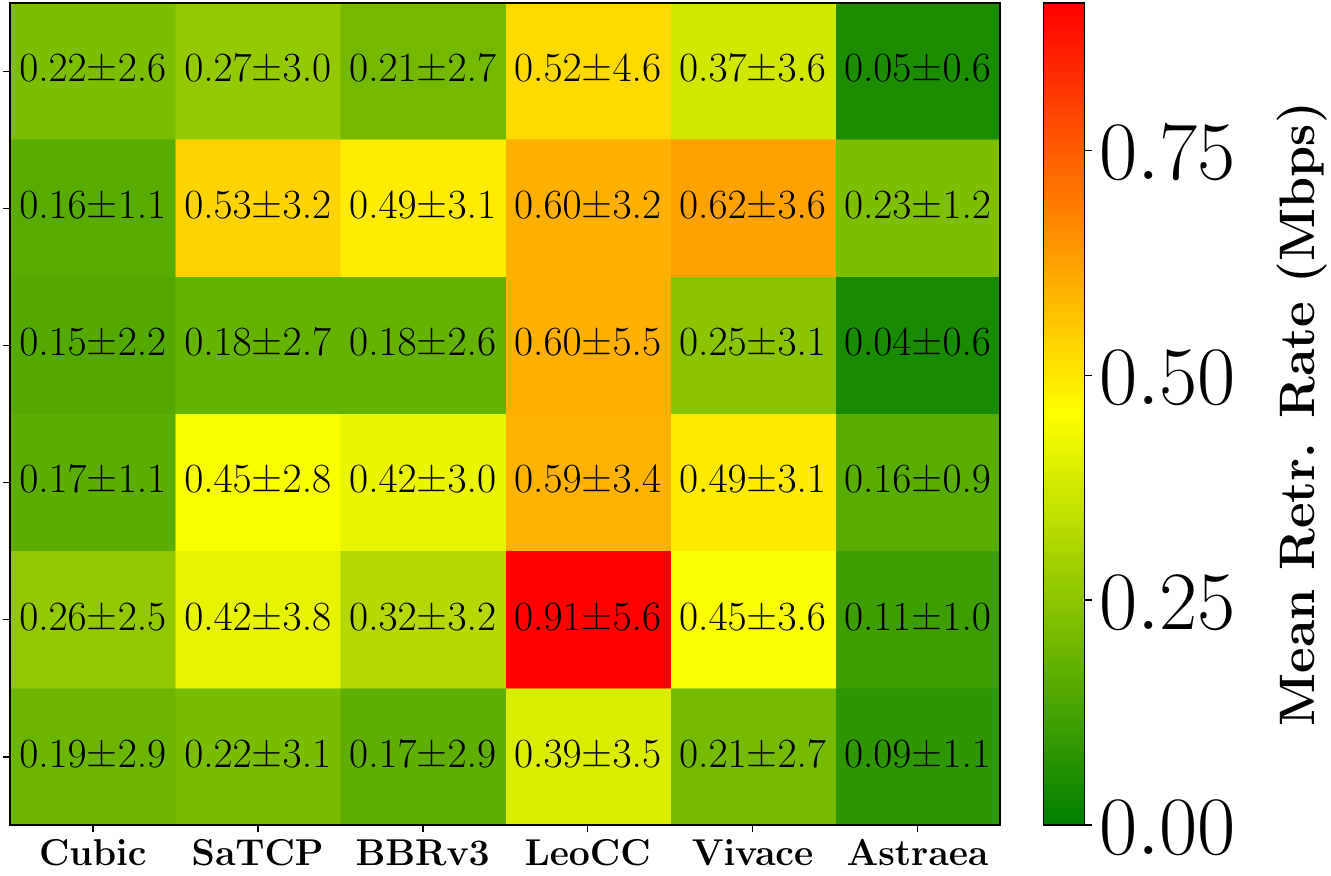}
        \vspace{-0.5cm}
        \caption{Retransmission rate}
        \label{fig:leoem_heatmap_retrans}
    \end{subfigure}
    \begin{subfigure}[t]{0.23\textwidth}
        \centering
        \includegraphics[width=\linewidth]
            {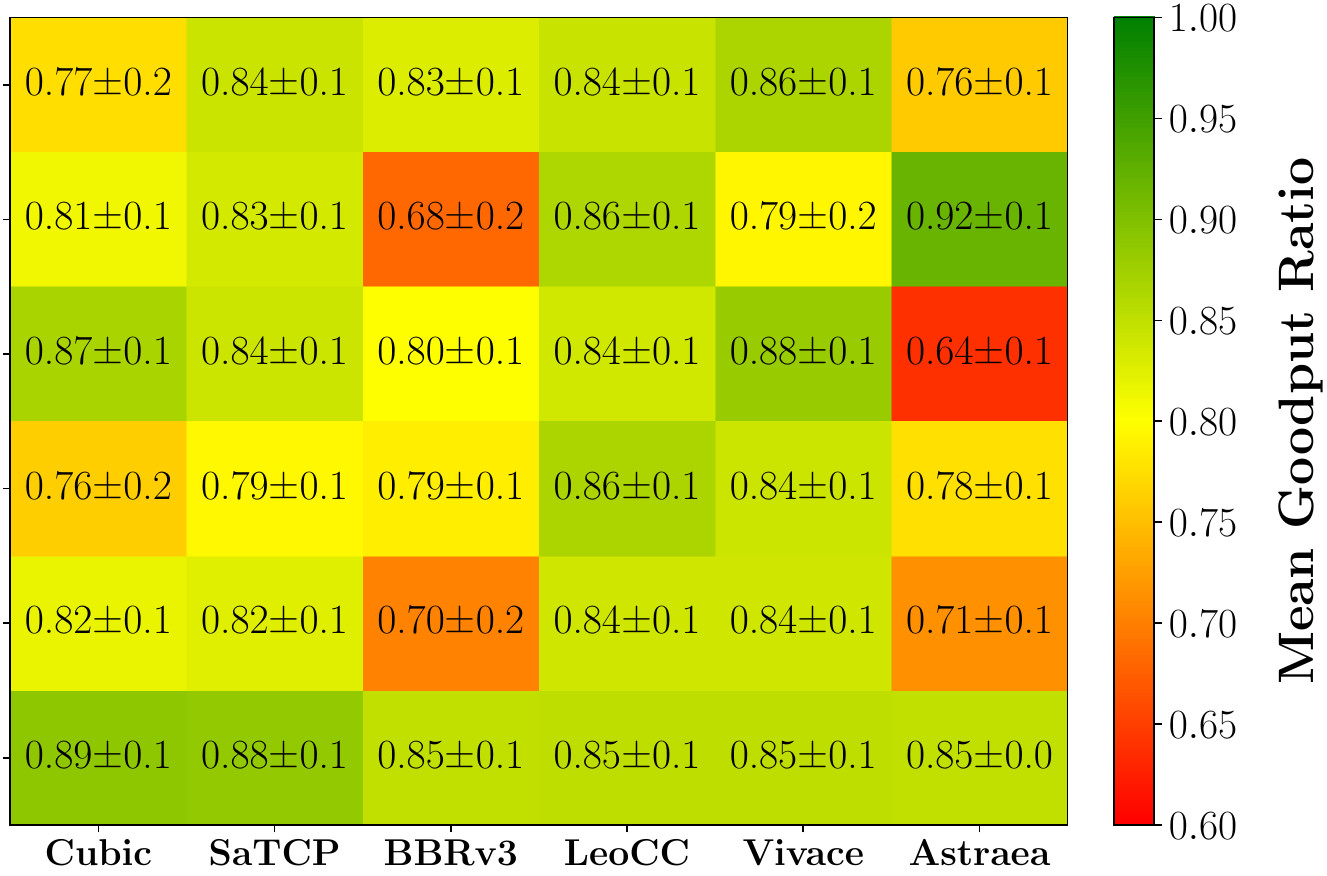}
        \vspace{-0.5cm}
        \caption{Goodput ratio}
        \label{fig:leoem_heatmap_fairness}
    \end{subfigure}
    \caption{LeoEM Experimentation with a single flow (a), (b) and (c), and two competing flows (d)}
    \label{fig:leoem_heatmaps}
    \vspace{-0.4cm}
\end{figure*}

\paragraph{\textbf{Performance Metrics and Data Collection}}
For Cubic and BBRv3, we used iPerf3 \cite{iperf} to generate traffic and collect goodput measurements. Astraea comes with its own bespoke applications, which we used to record goodput. For Vivace, we used the UDT API \cite{udt} to access all needed metrics. We collect additional metrics using \cite{ss}. Goodput measurements were performed at $1$-second intervals. All experiments were run five times, except Section \ref{sec:responsiveness}, which was run 50 times. We focus primarily on averaged metrics that capture efficiency, latency inflation, and fairness. 
Goodput is our main performance metric, as it directly reflects user-perceived performance. For fairness evaluation, we compute the goodput ratio, defined as the goodput of the lower-throughput flow divided by that of the higher-throughput flow; this metric is only applied in experiments involving two simultaneously active flows; when more flows are present, we use Jain's fairness index. To quantify delay inflation, we calculate the ratio of measured smoothed RTT to the base RTT.
\section{Experimentation with LeoEM}
\label{sec:LeoEM}

In this section, we use LeoEM to evaluate how the selected CC schemes perform on emulated LEO satellite network paths. Specifically, we set up data flows on paths built using ISLs or BP links between emulated user terminals deployed in London (LDN), San Diego (SD), Seattle (SEA), New York (NY) and Shanghai (SHA). Through this experiment, we observe how handovers and fluctuating RTTs impact CC performance. We carried out two sets of experiments: one measuring the goodput achieved in a single-flow setup, and another comparing fairness when two flows coexist on a shared LEO path.

\begin{figure*}[ht]
  \centering
  \includegraphics[width=0.9\textwidth]{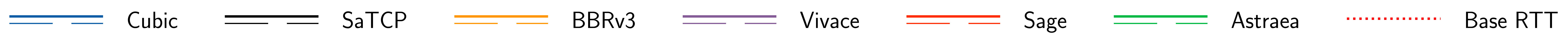}
  
  \begin{subfigure}[b]{0.32\textwidth}
    \centering
    \includegraphics[width=\textwidth]{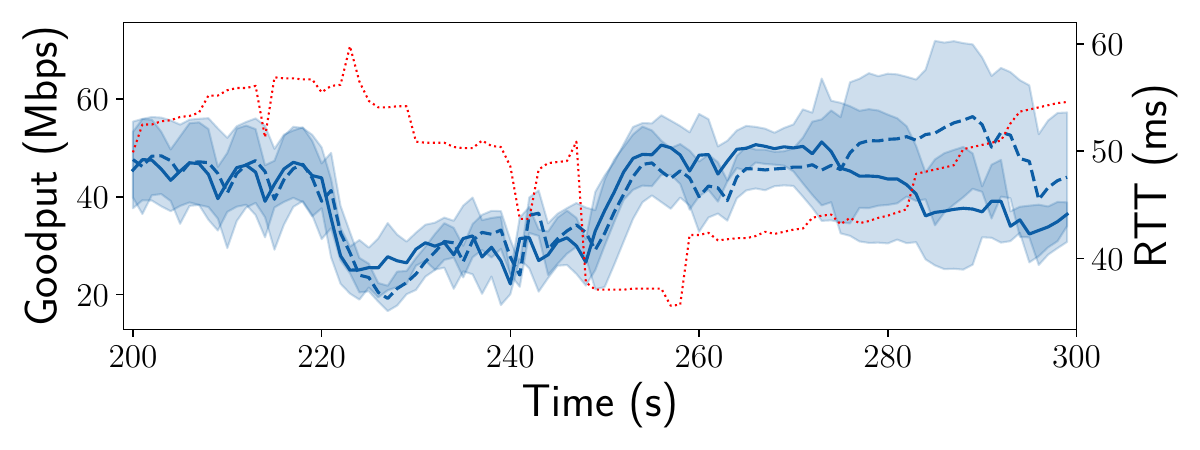}
    \vspace{-0.6cm}
    \caption{Cubic}
    \label{fig:cubic_leo}
  \end{subfigure}\hfill
  \begin{subfigure}[b]{0.32\textwidth}
    \centering
    \includegraphics[width=\textwidth]{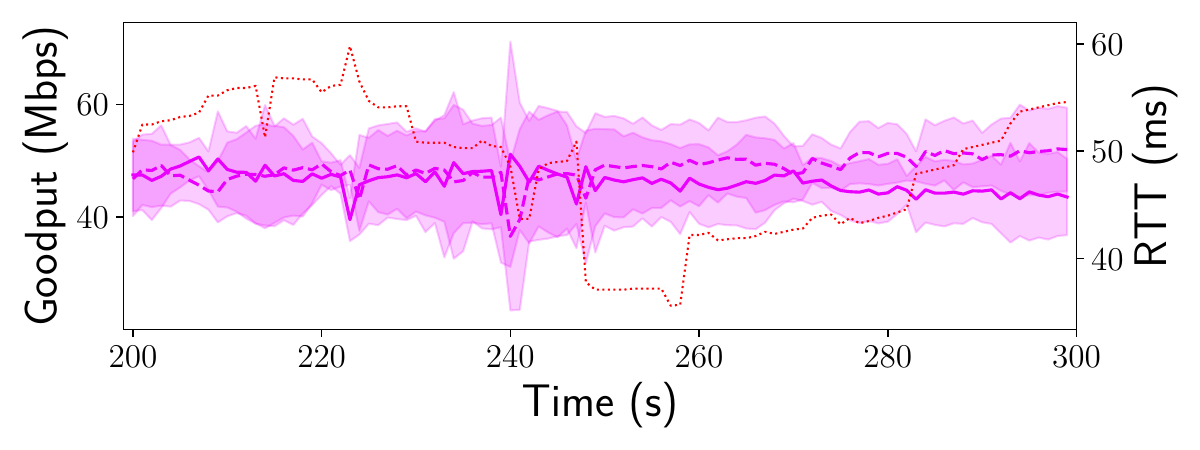}
    \vspace{-0.6cm}
    \caption{SaTCP}
    \label{fig:satcp_leo}
  \end{subfigure}\hfill
  \begin{subfigure}[b]{0.32\textwidth}
    \centering
    \includegraphics[width=\textwidth]{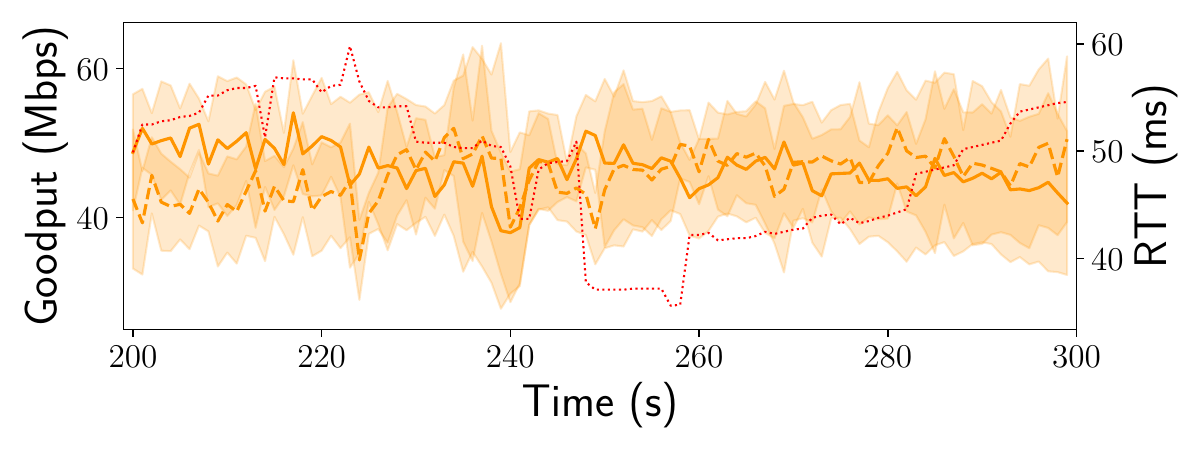}
    \vspace{-0.6cm}
    \caption{BBRv3}
    \label{fig:bbrv3_leo}
  \end{subfigure}

  \begin{subfigure}[b]{0.32\textwidth}
    \centering
    \includegraphics[width=\textwidth]{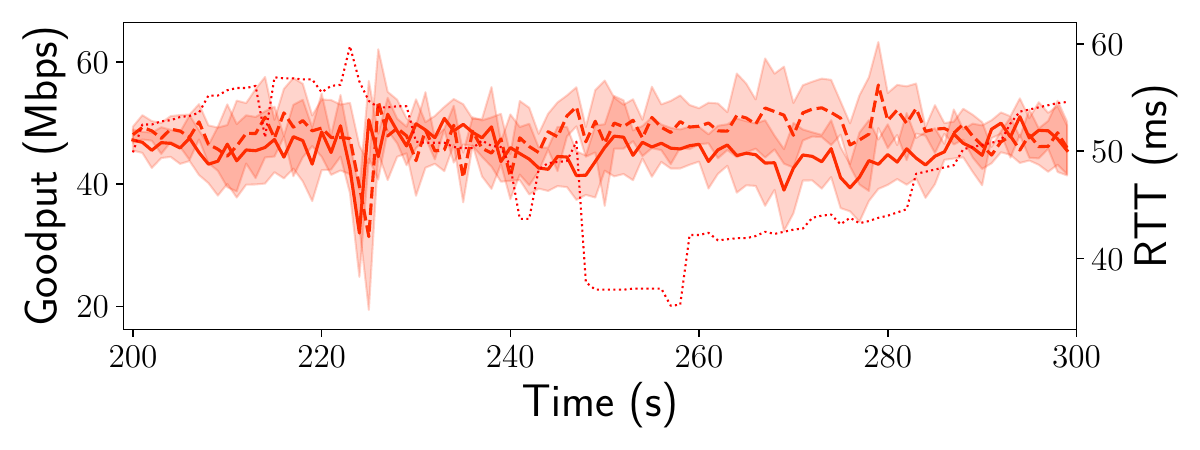}
    \vspace{-0.6cm}
    \caption{LeoCC}
    \label{fig:leocc_leo}
  \end{subfigure}\hfill
  \begin{subfigure}[b]{0.32\textwidth}
    \centering
    \includegraphics[width=\textwidth]{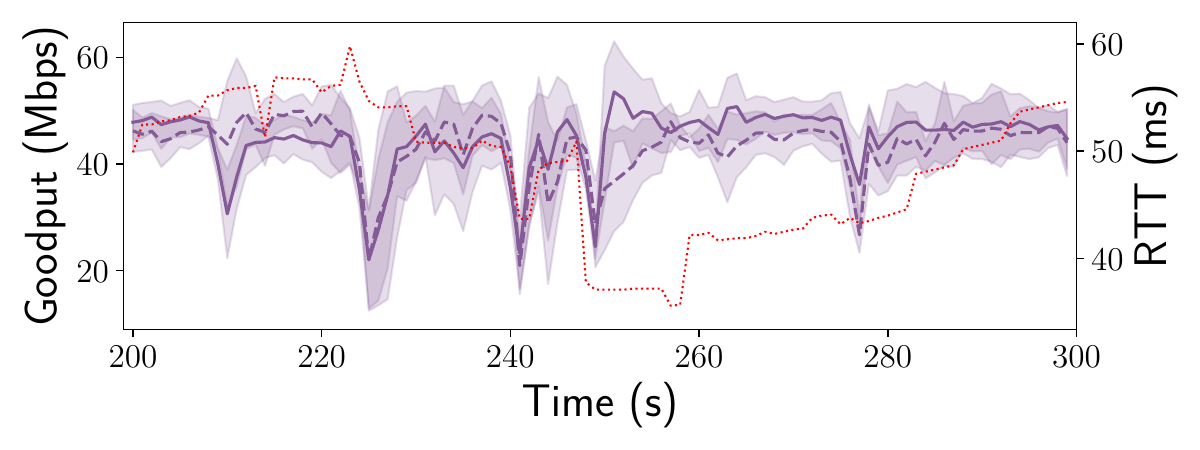}
    \vspace{-0.6cm}
    \caption{Vivace}
    \label{fig:vivace_leo}
  \end{subfigure}\hfill
  \begin{subfigure}[b]{0.32\textwidth}
    \centering
    \includegraphics[width=\textwidth]{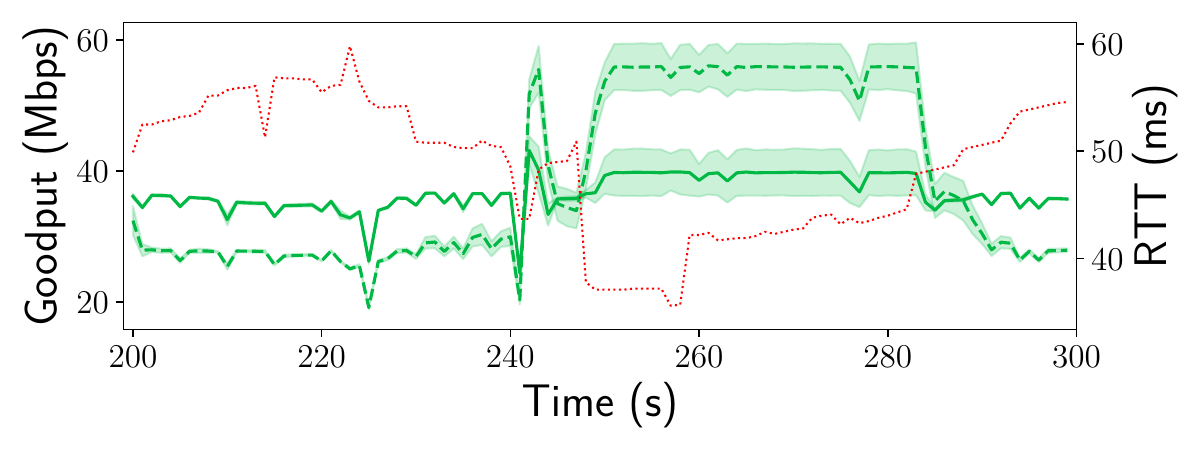}
    \vspace{-0.6cm}
    \caption{Astraea}
    \label{fig:astraea_leo}
  \end{subfigure}

  \caption{Goodput evolution of competing flows (solid/dashed lines) and base RTT of the SD to NY (BP) path (red dotted line).}
  \label{fig:leoem_fairness_evolution}
  \vspace{-0.4cm}
\end{figure*}

\subsection{Single-Flow Performance} 
\label{sec:LeoEM_Goodput}

For this experiment, we used the paths described above and ran a single flow from the respective user terminals for $300$ seconds. The buffer capacity of each switching device along a path is set to $1\times$ the BDP of that path. In Figures \ref{fig:leoem_heatmap_goodput} and \ref{fig:leoem_heatmap_delay}, we report the average goodput, and delay inflation, respectively. 

Cubic is sensitive to the frequency of handovers. For example, on the SD to NY path, Cubic is unable to capture all the available bandwidth, and this is because the frequency of handovers is high on this BP path. As a result, the associated packet loss causes frequent congestion window reductions at a rate that Cubic cannot recover from fast enough. SaTCP outperforms Cubic in interruption-prone scenarios by conserving its congestion window through interruptions. 

Although the loss-based approach of Cubic combined with the handover-aware SaTCP is effective in capturing the available bandwidth, it does so at the expense of a higher retransmission rate (seen in Figure \ref{fig:leoem_heatmap_retrans}) and substantial delay inflation, as observed in Figure \ref{fig:leoem_heatmap_delay}, reaching up to $2.14\times$ for Cubic (on a best case low interruption path) and $2.30\times$ for SaTCP. Consequently, while LEO links reduce end-to-end propagation delay, these large queueing delays inflate the RTT (just as they can in terrestrial networks), eroding or even eliminating the practical latency advantage that a LEO network would otherwise provide (Section \ref{intro}).

BBRv3 tracks the available bandwidth relatively well on all tested LEO paths (Figure \ref{fig:leoem_heatmap_goodput}), falling only slightly behind SaTCP, albeit with a much lower delay inflation. This under‐utilisation occurs whenever the base RTT jumps sharply and BBRv3 continues pacing at an outdated, lower BDP rate, leading to a transient reduction in its send rate. Recovery only happens after the $10$-second minimum RTT filter expires and a subsequent bandwidth probe refreshes its model, at which point it resumes sending to the path's bottleneck bandwidth. This behaviour mirrors observations in BBRv1 \cite{satcp, towards_a_deeper_understanding_of_bbrv1} and persists in BBRv3 due to its unchanged RTT filter length. BBRv3 does not significantly inflate the delay on highly dynamic BP (e.g., SEA to NY) or more stable ISL paths (e.g., SD to SHA), as shown in Figure \ref{fig:leoem_heatmap_delay}. 

Much like SaTCP, LeoCC manages to capture all available bandwidth, though with a much lower delay inflation on all evaluated LEO paths (Figure \ref{fig:leoem_heatmap_delay}). It extends BBRv1 by leveraging a companion user-space application that explicitly signals the sender whenever a path reconfiguration is detected, as well as additional bandwidth and RTT filters to tackle noisy observations \cite{leocc}. Upon receiving this signal, the sender briefly reduces its sending rate and resets the network model. LeoCC then immediately re-enters BBR STARTUP with a conservative initial congestion window of roughly half the current BDP. This triggers a rapid refresh of the bandwidth and RTT estimates after each reconfiguration, preventing the prolonged under-utilisation that can occur when BBR continues pacing using stale, pre-change measurements. However, LeoCC exhibits the highest retransmission rate among the evaluated protocols. This elevated retransmission rate is attributable to the startup phase of the reconfiguration. The exponential ramp-up of startup will overshoot the available capacity and wait for three consecutive RTTs before exiting, inducing packet loss and subsequent retransmissions, as it is repeated every time a reconfiguration takes place.

Similarly to BBRv3, Vivace does not capture all the available bandwidth. This is because when the base RTT is high and Vivace reduces its sending rate based on the utility observed when testing higher and lower sending rates, it takes time for it to recover. We discuss this in more detail in Section \ref{sec:responsiveness}. As shown in Figure \ref{fig:leoem_heatmap_delay}, Vivace's delay inflation is similar to BBRv3, and this is because Vivace continuously probes lower but also higher sending rates, with the latter driving latency inflation and retransmissions. 

Although most RL-based CC schemes have been shown to be dynamic and responsive \cite{orca, astraea, sage}, in a LEO context, we observe that Astraea performs very poorly in terms of bandwidth acquisition in all tested LEO satellite network paths. Due to severe bandwidth under-utilisation, Astraea does not inflate the paths' RTTs as queues are almost always empty, and the retransmission rate is mainly driven by the reconfigurations. Astraea is not trained in environments with changing base RTTs, although agents have seen a wide range of network conditions. Any changes in the base RTT will be interpreted during inference as a sign of congestion, and as such bandwidth exploration will be hindered. This limitation, also noted in \cite{lucapaper_systematic_rlcc}, suggests that the community should pursue alternative approaches, either by redesigning the reward formulation or by rethinking the training setup to better match LEO dynamics.

\begin{figure*}[t]
\centering
    \begin{minipage}[ht]{0.33\textwidth}
    \centering
    \subcaptionbox{\textbf{Responsiveness}. CDF of Goodput \label{fig:response_goodput_cdf}}{
    \includegraphics[width=\linewidth]{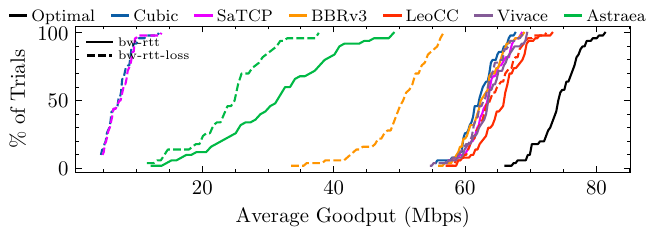}}
    \subcaptionbox{\textbf{Retransmissions}. CDF of Retransmissions \label{fig:response_retr_cdf}}{
    \includegraphics[width=\linewidth]{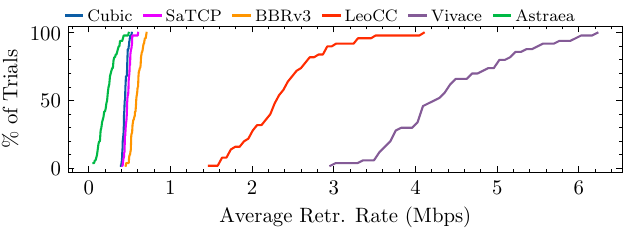}}
    \end{minipage}
    \hfill
    \begin{minipage}[ht]{0.66\textwidth}
    \centering
    \includegraphics[width=\linewidth]{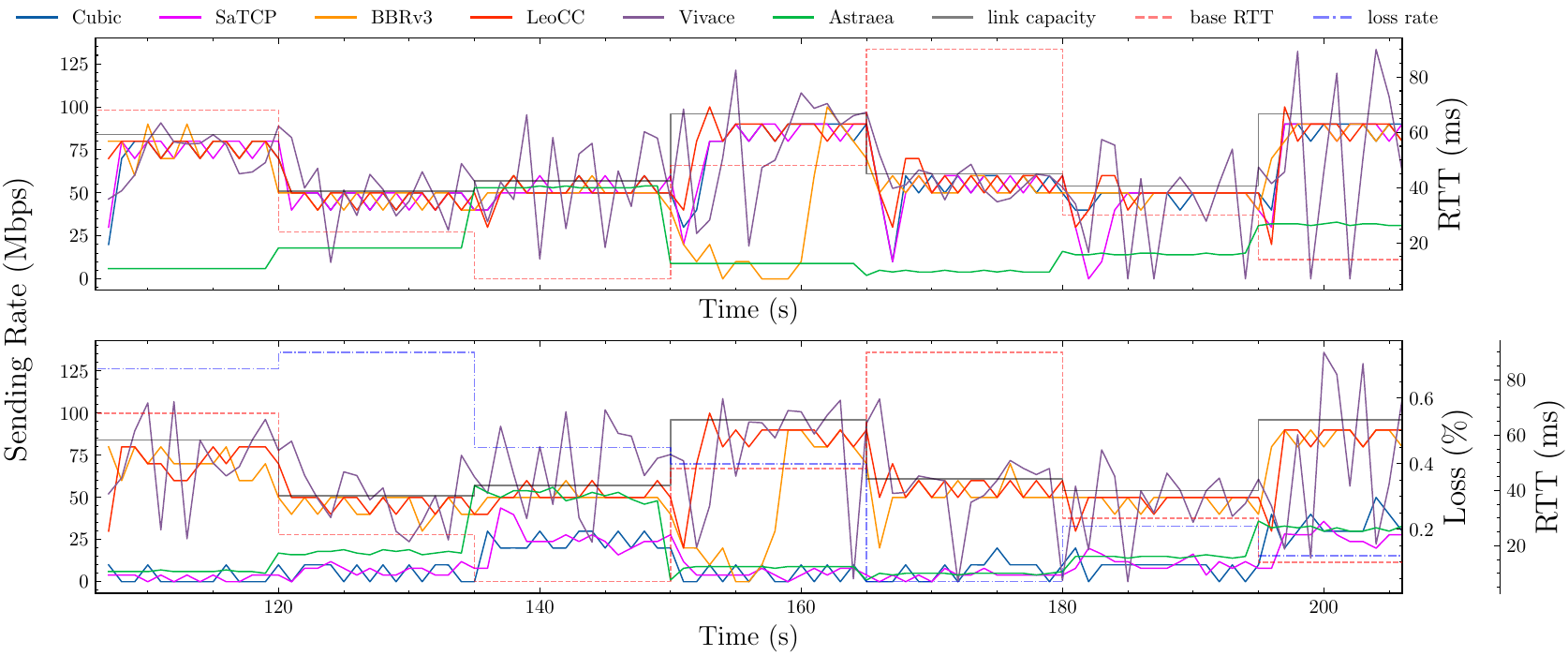}
    \subcaption{\textbf{Responsiveness}. Sending Rate in Time}
    \label{fig:response_send_rate}
    \end{minipage}
\vspace{-0.2cm}
\caption{\textbf{Responsiveness}. Sending rate over time (left) and cumulative distribution of goodput (right, repeated and stacked).}
\label{fig:response_combined_stacked_side}
\vspace{-0.3cm}
\end{figure*}

\subsection{Two flow fairness}
\label{sec:LeoEM_fairness}
In this section, we evaluate the ability of each CC scheme to maintain fairness under the dynamic conditions present in LEO satellite networks. In this experiment, we start the second flow $100$ seconds after the first flow and let both flows run concurrently for $300$ seconds. In Figure \ref{fig:leoem_heatmap_fairness}, we report the average goodput ratio of the lower throughput flow over the higher throughput flow. Measurements are from the last $100$ seconds to let flows converge. In Figure \ref{fig:leoem_fairness_evolution}, we show the goodput evolution for the two competing flows on the San Diego - New York path, for all tested CC schemes.

As illustrated in Figure \ref{fig:leoem_heatmap_fairness}, Cubic achieves high fairness on the most dynamic BP paths, indicated by goodput ratios closer to $1$. However, Cubic experiences periodic under-utilisation, which disrupts genuine contention between the two flows and can delay convergence to the fair operating point, since repeated drops reset or slow its progression toward equilibrium. This behaviour is evident on the SD - NY (BP) path (Figure \ref{fig:cubic_leo}) around the 220-second mark, where both flows appear to share bandwidth well largely because neither fully occupies the available capacity. In contrast, SaTCP’s handover-aware mechanism mitigates this under-utilisation while otherwise behaving like standard Cubic, preserving AIMD-like fairness.

BBRv3 generally maintains high fairness across all tested paths. In Figure \ref{fig:bbrv3_leo}, we observe that the convergence behaviour of BBRv3 is noisy, and this is the result of the $5$-second `probe RTT' phase, combined with our $1$-second sampling frequency. LeoCC maintains consistently high fairness in both high and low interruption paths (Figure \ref{fig:leoem_heatmap_fairness}), and the two flows share capacity with less variability (Figure \ref{fig:leocc_leo}). As noted in \cite{promisesofbbrv3}, BBRv1 is typically more fair when competing flows start simultaneously. Since LeoCC restarts (or re-enters) startup at each reconfiguration, it repeatedly brings flows into a more synchronised state, which likely contributes to the improved fairness we observe.

Vivace achieves very high fairness, particularly on stable ISL paths, as shown in Figure \ref{fig:leoem_heatmap_fairness}. The Vivace results confirm the theoretical results discussed in \cite{vivace_uspace} stating that when any number of senders share a bottleneck link and all senders share the Vivace utility function, the sending rates converge to a fair configuration \cite{vivace_uspace}. A closer look at the goodput evolution illustrated in Figure \ref{fig:vivace_leo} shows that the two flows have a noisy convergence profile. This is attributed to Vivace's high-frequency probing rate trials. Although Vivace is shown to be fair when the two flows experience the same base RTT, in Section \ref{sec:fairness_micro}, we show that it becomes very unfair when the competing flows experience different base RTTs. Astraea, the first RL approach to integrate fairness directly into its reward formulation, maintains mixed fairness across the paths tested, as seen in Figure \ref{fig:leoem_heatmap_fairness}. For all paths, Astraea suffers from under-utilisation, due to the changing base RTT. In the SD - SHA (ISL) path, the base RTT changes are less drastic, and flows achieve a goodput ratio more indicative of Astraea's reported fairness \cite{astraea}. As shown in Figure \ref{fig:astraea_leo}, the two flows are initially unfair to each other because they experience different base RTTs. Around the $250$ second mark, when the base RTTs converge to their lowest value, the flows compete more evenly and achieve fairer bandwidth sharing, we explore this further in Section \ref{sec:responsiveness}. In addition to fairness, Astraea’s reward formulation includes a stability component, which is effective in this scenario, as indicated by the little variance in goodput seen in Figure \ref{fig:astraea_leo}.

\section{Micro-benchmarks}
\label{sec:microbenchmarks}

In the previous section, we quantified the ability of each CC scheme to capture available bandwidth and examined how these schemes share bandwidth under highly dynamic conditions. Fairness and responsiveness alone do not capture the full picture; factors such as heterogeneous RTTs, multiple bottlenecks, and non-congestive loss (e.g., due to rain fade \cite{rain_fade}) are not captured in LeoEM. To address these concerns, we now turn to micro-benchmarks that enable us to isolate individual network scenarios, providing a more interpretable evaluation of the selected schemes, using the setup described in \ref{par:experimentalSetup}. We isolate important network scenarios, such as (1) rapid fluctuations in bandwidth, RTT variations and non‑congestive loss, (2) intra- and inter-RTT fairness, (3) multi-bottleneck scenarios and (4) how efficient each scheme is when sharing the bottleneck, with the added 15 second reconfiguration typical of Starlink.

\subsection{Responsiveness}
\label{sec:responsiveness}

In this section, we study how the selected CC approaches react to dynamically changing network conditions similar to the ones encountered in Section \ref{sec:LeoEM} (e.g. due to link re-configuration and routing changes, or transient hotspots). Through our micro-benchmarks, we can explore how CC approaches behave in the presence of changes in bandwidth, base RTT, and random loss rate, without the effects of lossy handovers. This is not possible with LeoEM. 

\noindent\textbf{Varying Bandwidth and Base RTT}. We experiment with a single $300$-second flow over an emulated LEO path. Every $15$ seconds we update the bottleneck bandwidth and the base RTT parameters by uniformly selecting values from the ranges $50$ - $100$Mbps and $10$ - $100$ms, values comparable to the paths chosen in Section \ref{sec:LeoEM}. We have opted for a 15-second change interval to match the rate at which Starlink user terminals perform link reconfigurations \cite{starlink_performance} with a reconfiguration length of $45$ - $120$ms, also used in Section \ref{sec:LeoEM}. The buffer capacity is set to $1\times$ the mean BDP of the emulated path. For each CC approach, we repeat the experiment $50$ times and calculate the cumulative distribution of average goodput, as in \cite{vivace_uspace}. In each run, all tested CC schemes experience the same bandwidth and base RTT variations, so we can directly compare their performance. The results of this experiment are shown in Figure \ref{fig:response_goodput_cdf}; the optimal goodput is shown with the black line. Figure \ref{fig:response_send_rate} (top) shows the evolution of the sending rate for all tested CC schemes. 

Cubic follows the available bandwidth despite changes in the baseline RTT estimate (solid blue line), but it achieves goodput slightly lower than BBRv3. The recurring 15-second interruptions repeatedly restart Cubic’s congestion window growth epoch, preventing the window from consistently reaching its peak before losses occur and resulting in reduced goodput. In comparison, SaTCP achieves better average goodput using its congestion window preserving mechanism, which allows it to maintain congestion window growth throughout the reconfiguration. 

BBRv3 tracks the available bandwidth well. We observed a similar reduction in BBRv3's goodput in the LeoEM experiment, in Section \ref{sec:LeoEM_Goodput}. This is attributed to its minimum RTT filter which, combined with its wall clock bandwidth probing approach, causes a slightly slower adaptation to bandwidth increases, as it must wait at most 3 seconds to start the `probe bandwidth' phase to explore newly available bandwidth. This can be clearly seen in Figure \ref{fig:response_send_rate} (top) at the 150-second mark, where BBRv3 takes around 12 seconds to adapt to the sharp increase in base RTT, during which it maintains a much lower sending rate than optimal. LeoCC accurately tracks the available bandwidth the best out of all protocols, as in the experiments in Section \ref{sec:LeoEM}, quickly adjusting to the changes in bandwidth regardless of the magnitude of the RTT jumps, due to the model of the network being reset. 

Vivace performs in line with the LeoEM experimentation (solid purple line). Vivace's responsiveness deficit is attributed to (1) its monitoring interval being based on the underlying RTT, (2) the bandwidth growth cap Vivace employs to filter out erroneous values, and (3) the periodic sending rate reversal to a much lower rate when utility is no longer increasing. Vivace ramps up its sending rate slowly because of the large RTT increase around the 190-second mark in Figure \ref{fig:response_send_rate}, and the characteristic sending-rate dips persist throughout the experiment. Astraea is unable to respond to frequent base RTT and bandwidth fluctuations, achieving around half the optimal performance (solid green line). In Figure \ref{fig:response_send_rate} (top), we observe that the sending rate of Astraea collapses when the base RTT is higher than the minimum observed RTT. Up to the 135-second mark, the base RTT is the lowest, and the Astraea flow captures all available bandwidth. Subsequently, Astraea only sees higher base RTT values than its recorded minimum value, and consequently, it is unable to capture much of the available bandwidth. The Astraea RL agent uses the current RTT divided by the minimum RTT seen so far as a congestion signal. Any time the measured base RTT exceeds its historical minimum (a typical occurrence on a LEO satellite network path), it is interpreted as queue build-up, which it has learnt to counteract by reducing its sending rate. Astraea has not seen varying RTT values during training episodes.

\begin{figure*}[!ht]
    \centering
    \includegraphics[width=0.75\textwidth]{figures/protocol_legend_squares.pdf}
    \begin{subfigure}[t]{0.32\textwidth}
        \includegraphics[width=\linewidth]{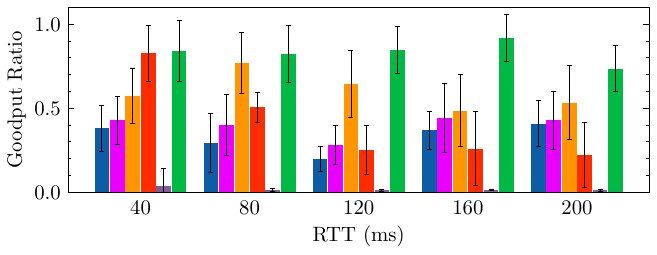}
        \vspace{-0.7cm}
        \caption{Buffer Size: $0.2\times$ BDP }
        \label{fig:interRTTMininet0.2}
    \end{subfigure}
    \begin{subfigure}[t]{0.32\textwidth}
        \includegraphics[width=\linewidth]{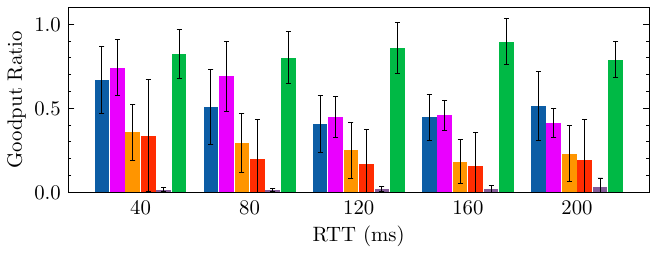}
        \vspace{-0.7cm}
        \caption{Buffer Size: $1\times$ BDP }
        \label{fig:interRTTMininet1}
    \end{subfigure}
    \begin{subfigure}[t]{0.32\textwidth}
        \includegraphics[width=\linewidth]{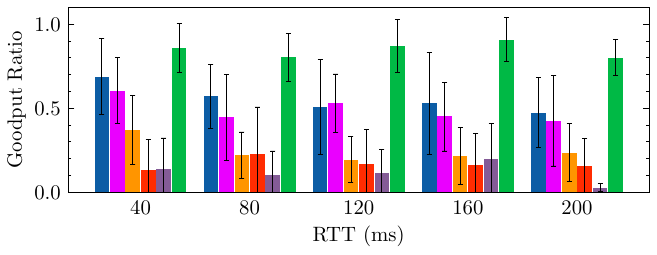}
        \vspace{-0.7cm}
        \caption{Buffer Size: $4\times$ BDP }
        \label{fig:interRTTMininet4}
    \end{subfigure}
    \caption{Goodput ratio of two competing flows on an emulated LEO satellite path. Starting flow has 20 ms RTT and the joining flow's RTT is on the x-axis. Queue size is set based on the BDP of the larger flow. }
    \vspace{-0.3cm}
    \label{fig:Inter-RTTFairness}
\end{figure*}

\noindent\textbf{Varying Bandwidth, Base RTT and Non-Congestive Loss}. Here, we add random packet loss to the setup above to emulate effects of harsh weather \cite{rain_fade}. We do so by uniformly selecting a random loss probability from the range $0$\% - $1$\%. We plot the cumulative goodput distribution in dashed lines in Figure \ref{fig:response_goodput_cdf} and the evolution of the sending rate in Figure \ref{fig:response_send_rate} (bottom).

Cubic is known to perform very poorly in the presence of non-congestive loss, being a loss-based protocol; this is clearly shown in Figure \ref{fig:response_goodput_cdf} (dashed blue line). In Figure \ref{fig:response_send_rate}, we observe that Cubic's sending rate collapses due to the recurring non-congestive loss. 
BBRv3 struggles to track the bandwidth it would otherwise, while it experiences non-congestive loss (dashed orange line). In this setup, BBRv3 allows bandwidth probing only while the per-round-trip loss rate stays below its configured loss threshold (default 2\%) \cite{bbrv3}. When loss exceeds that threshold, BBRv3 treats the current bytes in-flight volume as too aggressive, pausing further bandwidth exploration. This is clearly seen at 180 seconds (Figure \ref{fig:response_send_rate} (bottom)), where BBRv3 is unable to explore the available bandwidth due to the higher non-congestive loss. This result highlights the drawbacks of the congestion window freeze approach used in \cite{starquic, satcp}, which is that it does not address the non-congestive loss that LEO paths may experience. In contrast, LeoCC uses BBRv1 as the basis for its implementation. Since BBRv1 largely ignores loss and LeoCC does not rely on packet loss as a congestion signal, LeoCC is resistant to non-congestive loss, tracking the available bandwidth similarly well to the non-loss setup. Vivace tracks the available bandwidth well in the presence of non-congestive loss (dashed purple line), as \cite{vivace_uspace} suggests. Astraea (dashed green line) is resistant to non-congestive loss (similar to the findings of \cite{lucapaper_systematic_rlcc}), tracking the available bandwidth very closely compared to the lossless setup, albeit still exhibiting poor performance overall. 

\begin{figure}
    \centering
    \includegraphics[scale=0.8]{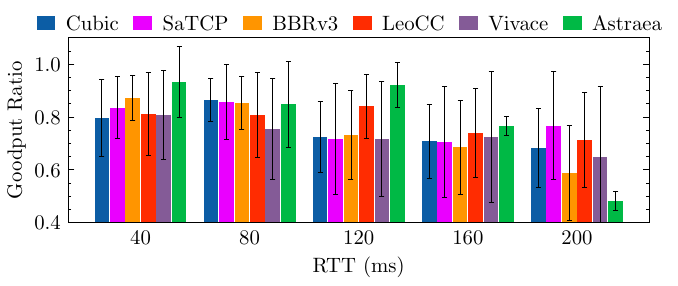}
    \vspace{-0.5cm}
    \caption{Goodput ratio of two competing flows on an emulated LEO satellite path, both experiencing the same RTT. }
    \label{fig:intra_rtt}
    \vspace{-0.4cm}
\end{figure}

\subsection{Fairness}
\label{sec:fairness_micro}
Fairness is a key consideration when evaluating CC schemes for use in LEO satellite networks. In this section, we look at both intra- and inter-RTT fairness; i.e., fairness when all competing flows experience the same base RTT and different base RTTs, respectively. Micro-benchmarks are invaluable here because LeoEM can only emulate a single path and respective user terminals, and therefore it is not possible to examine inter-RTT fairness at all. Moreover, we want to examine fairness without the added effects of RTT variability and handovers present in LeoEM, to ensure we evaluate the fundamental fairness traits of the selected CC schemes.

\begin{figure*}[t]
    \centering
    \includegraphics[width=0.75\linewidth]{figures/protocol_legend_squares.pdf}\par\medskip
    \vspace{-0.3cm}
    \begin{minipage}[t]{0.66\textwidth}
        \centering
        \begin{subfigure}[t]{0.49\linewidth}
            \centering
            \includegraphics[width=\linewidth]{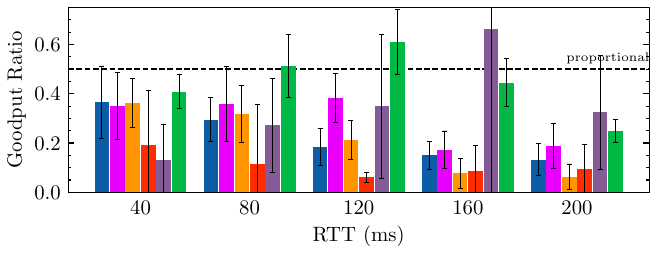}
            \caption{Bottlenecks: $3$}
            \label{fig:parkingLot_3hops}
        \end{subfigure}\hfill
        \begin{subfigure}[t]{0.49\linewidth}
            \centering
            \includegraphics[width=\linewidth]{figures/Parking_Lot/goodput_ratio_between_hops_goodput_flows3.pdf}
            \caption{Bottlenecks: $6$}
            \label{fig:parkingLot_6hops}
        \end{subfigure}
        \caption{\textbf{Multi-bottleneck fairness:} Goodput ratio of the multi-bottlenecked flow (traversing all bottlenecks) to the best single bottleneck flow. } 
        \label{fig:parking_lot_fairness}
    \end{minipage}\hfill
    \begin{minipage}[t]{0.33\textwidth}
        \centering
        \includegraphics[width=\linewidth]{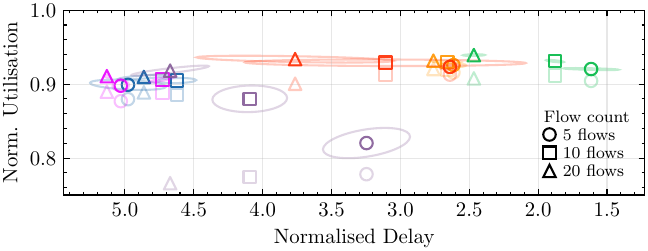}
        \caption{\textbf{Efficiency:} Normalised utilisation vs normalised delay of multiple flows sharing the bottleneck.}
        \label{fig:efficiency}
    \end{minipage}
\vspace{-0.3cm}
\end{figure*}

\paragraph{\textbf{Inter-RTT fairness}} 
In LEO satellite networks, it is common for flows with significantly different base RTT values to share portions of their paths. In this experiment, we evaluate how the selected CC schemes perform under such conditions. Note that LeoEM does not support multiple simultaneous paths, making this micro-benchmark essential for evaluating inter-RTT fairness. We repeat the previous experiment, but fix the base RTT of one flow to $20$ms and vary the RTT of the joining flow, shown on the x-axis in Figure \ref{fig:Inter-RTTFairness}. We set the buffer capacity to the BDP of the joining flow.

Cubic is known to be RTT-unfair \cite{cubic}, and this behaviour is reflected in our experiments across all tested RTT values. In many cases, we observe goodput ratios of roughly $0.5$, indicating that one flow attains about twice the bandwidth of the competing flow. Since SaTCP builds on Cubic, it inherits this inter-RTT unfairness. The fairness profile of BBRv3 is heavily influenced by the available buffer capacity. With the queue capacity set to $1\times$ and $4\times$ the BDP (Figures \ref{fig:interRTTMininet1} and \ref{fig:interRTTMininet4}), the flow with the larger base RTT maintains a higher volume of in-flight bytes, leading to progressively unfair allocations as the base RTT disparity increases. 

Our experimental findings show that LeoCC exhibits the same inter-RTT unfairness of BBRv1, performing worse than BBRv3 across most of our parameter range. Notably, these results diverge from LeoCC’s own evaluation, which reports strong inter-RTT fairness \cite{leocc}. In our experiments, fairness remains broadly BBR-like and is more sensitive to transient dynamics. The slightly better inter-RTT fairness of BBRv3 compared to LeoCC is because the pacing gain increases from 0.75 to 0.91 during the bandwidth `probe down' phase \cite{bbrv3}, making the existing flow yield less of its bandwidth to the joining flow. When the buffer is very small (Figure \ref{fig:interRTTMininet0.2}), the higher RTT flow cannot claim a substantially higher buffer occupancy than the $20$ms flow, resulting in a more fair allocation than when the buffer capacity is $1\times$ the path's BDP. 

Vivace exhibits the weakest performance of all tested CC schemes. This result highlights that the fairness theorem, and the accompanying proofs \cite{vivace_uspace}, assume that all senders share an identical base RTT in order to achieve a fair bandwidth allocation. Looking at individual runs, the lower RTT flow always dominates the other flow. This is due to the RTT-based probing that Vivace does, probing both increased and decreased sending rates to maximise utility. The lower RTT flow will subsequently conduct its probing at a much faster rate than the higher RTT flows, causing unfairness. Across the entire range of tested RTTs, Astraea consistently maintains the highest performance, substantially outperforming all schemes, as shown in Figure \ref{fig:Inter-RTTFairness}. Astraea's multi-agent training approach, in combination with its fixed and RTT-independent monitoring time interval, is very effective. Interestingly, Astraea maintains its fairness properties even when one of the present flows is outside its training parameters.

\paragraph{\textbf{Intra-RTT fairness}}

In this experiment, the first flow runs for 200 seconds, with the second flow starting 50 seconds into the experiment. In Figure \ref{fig:intra_rtt}, we plot goodput ratios averaged over the final 100 seconds of each experiment. The base RTT experienced by both flows is varied (and shown on the x‑axis), while the buffer capacity is set to $1\times$ the BDP. In addition, every 15 seconds, we will emulate a link reconfiguration. 

Cubic and SaTCP perform very similar to each other when the base RTT remains consistent across the tested RTTs. BBRv3 shows similar performance to the LeoEM experiments, specifically the cases where under-utilisation was low, across all tested base RTT values, which is consistent with the findings of \cite{promisesofbbrv3}, where BBR is less fair when the flows do not start simultaneously. LeoCC achieves high fairness maintained throughout the periodic interruptions. Vivace achieves a fairness similar to that observed in Section \ref{sec:LeoEM_fairness}. We observe a higher variance, seen in Figure \ref{fig:intra_rtt}, indicating a noisy fairness profile. Astraea consistently achieves very high fairness, except for higher delays (see Figure \ref{fig:intra_rtt}), when the two flows fall outside Astraea’s training parameter range of $10$ - $140$ms. Beyond its training parameters, Astraea is unable to generalise its fairness properties.

\paragraph{\textbf{Multi-bottleneck fairness}} 
\label{sec:parking_lot}
The mesh-like topology of LEO satellite networks naturally leads to situations where flows encounter congestion and compete for bandwidth with other flows at multiple bottlenecks. This multi-bottleneck scenario also enables us to assess how different CC schemes perform with respect to max-min and proportional fairness. With max‐min fairness, bandwidth is distributed equally among all flows, while proportional fairness implies higher aggregate bandwidth utilisation. We emulate a multi-bottleneck LEO satellite network scenario using a parking lot topology (with $3$ and $6$ bottlenecks) and concurrently start flows that cross one of those bottlenecks each. Another flow that crosses all bottlenecks starts 50 seconds into the experiment. All flows experience the same base RTT. We plot the average ratio of the goodput of the multi-bottleneck flow over the highest goodput achieved amongst the single-bottleneck flows in Figure \ref{fig:parking_lot_fairness}. We look at the final 100 seconds of each run. Buffer capacity is sized to $1\times$ the BDP. 

Figures \ref{fig:parkingLot_3hops} and \ref{fig:parkingLot_6hops} illustrate the ratios for the three-bottleneck and six-bottleneck setups, respectively. We clearly observe that Cubic yields a fairness profile that is close to proportional fairness. Cubic, and by extension SaTCP, relies on packet loss as its congestion signal, and therefore, the multi-bottleneck flow will suffer packet loss at a higher rate. This, in turn, triggers more frequent multiplicative reductions of the congestion window than those experienced by single-bottleneck flows. This is known to happen with AIMD CC schemes \cite{fairnesscrowcroft}. Similarly, the rate-based convergence of BBRv3 achieves proportional fairness, since the single bottleneck flows will probe $3\times$ (Figure \ref{fig:parkingLot_3hops}) or $6\times$ (Figure \ref{fig:parkingLot_6hops}) more than the multi-bottleneck flow, having a higher chance to claim more bandwidth. LeoCC achieves proportional fairness in both three and six bottleneck scenario. The reconfiguration helps LeoCC to achieve the proportional fairness seen in Figure \ref{fig:parkingLot_6hops}, though like previous experiments, LeoCC also exhibits high variance. 
In these scenarios, the inter-RTT effects have minimal influence on the BBRv3 or LeoCC network models. Even though they do not share the same links, the single-bottleneck flows can still align their probe-RTT timing via the multi-bottleneck flow. In the three bottleneck scenario (Figure \ref{fig:parkingLot_3hops}), the multi-bottleneck Vivace flow struggles to sustain its sending rate, indicated by the ratio being under the proportional fairness line. We see in Figure \ref{fig:parkingLot_6hops}, that when we increase the number of bottlenecks, the multi-bottleneck flow collapses almost completely. This is likely the result of the multi-bottleneck flow experiencing a higher RTT causing unfairness similar to what we explored in Section \ref{sec:fairness_micro}. Astraea has not been trained using multi-bottleneck environments. When experiencing three bottlenecks (Figure \ref{fig:parkingLot_3hops}), Astraea achieves proportional fairness while operating within its range of delay training parameters. When we increase the number of bottlenecks to six, the multi-bottleneck Astraea flow does not capture much bandwidth at all for RTT values above $120$ms. A likely cause of this is the high delay the multi-bottleneck flow experiences as its packets must pass through six queues to reach their destination, adding up to an RTT value that is well beyond what Astraea has been trained with. 

\subsection{Efficiency}
\label{sec:efficiency}
In this section, we focus on how efficiently the different CC approaches utilise the underlying network resources, while multiple ($5$, $10$ and $20$) flows share the bottleneck. Flows are scheduled in evenly spaced intervals within the first 100 seconds, each lasting $150$ seconds. We measure aggregate goodput normalised by the capacity of the bottleneck, and average flow latency normalised by the path's base RTT for the $100$-$150$ second interval. We set the bottleneck bandwidth to $100$Mbps, the RTT to $50$ms and the buffer capacity to $5\times$ the BDP. In Figure \ref{fig:efficiency}, we plot mean values and $1$-$\sigma$ ellipses for the $50$ seconds where all flows co-exist. We use shaded marks to denote the normalised aggregate throughput after subtracting the normalised retransmission rate.

Cubic and SaTCP will achieve high utilisation at the cost of maximum delay inflation, for all tested concurrent flows ($5$, $10$ and $20$ flows) shown in Figure \ref{fig:efficiency}. The delay inflation is at its peak regardless of the number of flows sharing the bottleneck. The retransmissions are primarily driven by the link breakages that occur every 15 seconds. In our setting, BBRv3 achieves high utilisation, and delay inflation remains clustered near $2.5\times$ (seen in Figure \ref{fig:efficiency}) because each BBRv3 flow converges on essentially the same estimate of the aggregate bottleneck capacity, and BBR's model then targets roughly $2\times$BDP in flight. This aligns with prior findings in \cite{bbrv3wired} that all BBR variants tend to overestimate bottleneck bandwidth, which drives the queue build-up and sustained RTT inflation. Similarly, LeoCC matches BBRv3 in overall delay inflation (Figure \ref{fig:efficiency}), but its repeated startup contributes to more variable RTT inflation.

As described in Section \ref{sec:responsiveness}, Vivace exhibits periodic, brief drops in sending rate (Figure \ref{fig:vivace_leo}) when the probing trails reverse direction. As the number of flows increases, overall utilisation rises because these periodic reductions are increasingly filled by the other flows. Astraea, by contrast, maintains the smallest persistent queue among all schemes, and its size scales in proportion to the number of flows sharing the bottleneck. This is because each Astraea flow maintains a small per-flow delay inflation budget to enable safe bandwidth exploration by the agents during training. Although minimal for a few flows, these per-flow budgets add up as the number of simultaneous flows grows (e.g., $20$ flows shown in Figure \ref{fig:efficiency}), increasing the aggregate standing queue and thus overall RTT inflation.

\section{Conclusion}
\label{conclusion}

In this paper, we evaluated Cubic, SaTCP, BBRv3, LeoCC, Vivace, and Astraea using LeoEM LEO emulations and targeted micro-benchmarks. We find that (1) interruption-aware enhancement designs (SaTCP/LeoCC) improve bandwidth acquisition under frequent reconfigurations, but still inherit the same design trade-offs of the base schemes; (2) loss-based control (Cubic) is highly sensitive to interruption frequency and non-congestive loss, and tends to inflate delay when it succeeds in filling the pipe; (3) BBRv3 generally sustains high utilisation with moderate delay inflation, but remains limited in its responsiveness to sharp base-RTT shifts and becomes conservative under non-congestive loss; (4) Vivace is fair when flows share the same baseline RTT, yet its RTT-coupled probing leads to sluggish recovery after large RTT jumps and severe unfairness under RTT heterogeneity; and (5) Astraea is resilient to non-congestive loss and achieves strong fairness under RTT asymmetry and in several multi-bottleneck cases, but underutilises bandwidth when base RTT varies.

\bibliographystyle{IEEEtran}
\bibliography{references}

@INPROCEEDINGS{lucapaper_systematic_rlcc,
  author={Giacomoni, Luca and Parisis, George},
  booktitle={Proc of IEEE INFOCOM}, 
  year={2024},
  title={Reinforcement Learning-based Congestion Control: A Systematic Evaluation of Fairness, Efficiency and Responsiveness}, 
  }

@inproceedings{delay_not_an_option,
    author = {Handley, Mark},
    title = {Delay is Not an Option: Low Latency Routing in Space},
    year = {2018},
    booktitle = {Proc. of HotNets}
}

@inproceedings{towards_a_deeper_understanding_of_bbrv1,
    author={Scholz, Dominik and others},
    booktitle={Proc. of IFIP Networking}, 
    title={{Towards a Deeper Understanding of TCP BBR Congestion Control}}, 
    year={2018}
}

@inproceedings{21st_century_space_race,
    author = {Bhattacherjee, Debopam and others},
    title = {Gearing up for the 21st Century Space Race},
    year = {2018},
    booktitle = {Proc. of HotNets},
}

@inproceedings{network_topology_design_at_27000kmh,
    author={Bhattacherjee, Debopam and Singla, Ankit},
    title={Network Topology Design at 27,000 Km/Hour},
    year={2019},
    booktitle={Proc. of CoNEXT},
}

@inproceedings{starlink_performance,
    author = {Mohan, Nitinder and others},
    title = {{A Multifaceted Look at Starlink Performance}},
    year = {2024},
    booktitle = {Proc of the ACM Web Conference},
}

@inproceedings{leoroutingchallanges,
  author={Dai, Shiyue and Rui, LanLan and Chen, Shiyou and Qiu, Xuesong},
  booktitle={Proc. of IFIP/IEEE IM}, 
  title={{A Distributed Congestion Control Routing Protocol Based on Traffic Classification in LEO Satellite Networks}}, 
  year={2021}
}

@inproceedings{promisesofbbrv3,
    author = {Zeynali, Danesh and Weyulu, Emilia N. and Fathalli, Seifeddine and Chandrasekaran, Balakrishnan and Feldmann, Anja},
    title = {{Promises and Potential of BBRv3}},
    year = {2024},
    booktitle = {Proc. of PAM},
}

@article{fairnesscrowcroft,
author = {Crowcroft, Jon and Oechslin, Philippe},
title = {{Differentiated end-to-end Internet services using a weighted proportional fair sharing TCP}},
year = {1998},
journal = {SIGCOMM Comput. Commun. Rev.},
}

@article{li2007experimental,
  title={Experimental evaluation of {TCP} protocols for high-speed networks},
  author={Li, Yee-Ting and Leith, Douglas and Shorten, Robert N},
  journal={IEEE/ACM Transactions on Networking},
  volume={15},
  number={5},
  year={2007},
}

@inproceedings{astraea,
    author = {Liao, Xudong and Tian, Han and Zeng, Chaoliang and Wan, Xinchen and Chen, Kai},
    title = {Astraea: Towards Fair and Efficient Learning-based Congestion Control},
    year = {2024},
    booktitle = {Proc. of EuroSys},
}

@article{bbrv1,
    title	= {{BBR}: Congestion-Based Congestion Control},
    author	= {Neal Cardwell and Yuchung Cheng and C. Stephen Gunn and Soheil Hassas Yeganeh and Van Jacobson},
    year	= {2016},
    journal	= {ACM Queue},
}

@article{bbrv3wired,
title = {Evaluating TCP BBRv3 performance in wired broadband networks},
journal = {Computer Communications},
year = {2024},
author = {Jose Gomez and Elie F. Kfoury and Jorge Crichigno and Gautam Srivastava},


}

@misc{bbrv3,
  title={{BBRv3: Algorithm Bug Fixes and Public Internet Deployment}},
  author={Neal Cardwell and others},
  URL={https://datatracker.ietf.org/meeting/117/materials/slides-117-ccwg-bbrv3-algorithm-bug-fixes-and-public-internet-deployment-00},
  year={2023}
}

@article{cubic,
author = {Ha, Sangtae and Rhee, Injong and Xu, Lisong},
title = {{CUBIC: a new TCP-friendly high-speed TCP variant}},
year = {2008},
journal = {SIGOPS Oper. Syst. Rev.},
pages = {64–74},
numpages = {11}
}

@inproceedings{orca,
    author = {Abbasloo, Soheil and others},
    title = {{Classic Meets Modern: a Pragmatic Learning-Based Congestion Control for the Internet}},
    booktitle = {Proc. of ACM SIGCOMM},
    year = {2020},
}

@inproceedings{sage,
  title={{Computers Can Learn from the Heuristic Designs and Master Internet Congestion Control}},
  author={Yen, Chen-Yu and Abbasloo, Soheil and Chao, H Jonathan},
  booktitle={Proc. of ACM SIGCOMM},
  year={2023}
}

@inproceedings{satcp,
    title     ={{SaTCP: Link-Layer Informed TCP Adaptation for Highly Dynamic LEO Satellite Networks}},
    author    ={Cao, Xuyang and Zhang, Xinyu},
    year      ={2023},
    booktitle ={Proc. of IEEE INFOCOM},
}

@inproceedings {vivace_uspace,
    author = {Mo Dong and others},
    title = {{PCC Vivace: Online-Learning Congestion Control}},
    booktitle = {Proc. of USENIX NSDI},
    year = {2018}
}

@ARTICLE{rain_fade,
  author={Liu, Weiwen and Michelson, David G.},
  journal={IEEE Transactions on Vehicular Technology}, 
  title={{Fade Slope Analysis of Ka-Band Earth-LEO Satellite Links Using a Synthetic Rain Field Model}}, 
  year={2009},
  volume={58},
  number={8},
}

@misc{iPerf,
    title={iPerf},
    URL={https://iperf.fr/}
}

@misc{mininet, 
    title={Mininet},
    URL={http://mininet.org/}
}

@misc{tc,
    series =    {Linux tc},
    url =       {https://man7.org/linux/man-pages/man8/tc.8.html},
}

@inproceedings{hypatia,
    author = {Kassing, Simon and Bhattacherjee, Debopam and \'{A}guas, Andr\'{e} Baptista and Saethre, Jens Eirik and Singla, Ankit},
    title = {{Exploring the ``Internet from Space'' with Hypatia}},
    year = {2020},
    booktitle = {Proc. of ACM IMC},
}

@misc{bbrthreekernel, 
    title={{BBR3 enabled kernel}},
    URL={https://github.com/google/bbr/tree/v3}
}

@misc{ss, 
    title={ss(8) — Linux manual page},
    URL={https://man7.org/linux/man-pages/man8/ss.8.html}
}

@misc{udt, 
    title={UDT Documentation},
    URL={https://udt.sourceforge.io/udt3/}
}

@inproceedings{hock_experimental_2017,
  title = {Experimental Evaluation of {{BBR}} Congestion Control},
  booktitle = {Proc of ICNP},
  author = {Hock, Mario and Bless, Roland and Zitterbart, Martina},
  year = {2017},
}

@misc{valentine_developing_2021,
      title={{Developing and experimenting with LEO satellite constellations in OMNeT++}}, 
      author={Aiden Valentine and George Parisis},
      year={2021},
      archivePrefix={arXiv},
      primaryClass={cs.NI},
      url={https://arxiv.org/abs/2109.12046}, 
}

@inproceedings{ma_network_2023,
  author={Ma, Sami and Chou, Yi Ching and Zhao, Haoyuan and Chen, Long and Ma, Xiaoqiang and Liu, Jiangchuan},
  booktitle={Proc of IEEE INFOCOM}, 
  title={{Network Characteristics of LEO Satellite Constellations: A Starlink-Based Measurement from End Users}}, 
  year={2023},
}

@article{vegas,
    author = {Brakmo, Lawrence S. and O'Malley, Sean W. and Peterson, Larry L.},
    title = {TCP Vegas: new techniques for congestion detection and avoidance},
    year = {1994},
    issue_date = {Oct. 1994},
    publisher = {Association for Computing Machinery},
    url = {https://doi.org/10.1145/190809.190317},
    journal = {SIGCOMM Comput. Commun. Rev.},
}

@inproceedings{starquic,
    author = {Kamel, Victor and Zhao, Jinwei and Li, Daoping and Pan, Jianping},
    title = {{StarQUIC: Tuning Congestion Control Algorithms for QUIC over LEO Satellite Networks}},
    year = {2024},
    booktitle = {Proc of LEO-NET},
}

@inproceedings{hypatia_experiments,
    author = {Barbosa, George and Theeranantachai, Sirapop and Zhang, Beichuan and Zhang, Lixia},
    title = {{A Comparative Evaluation of TCP Congestion Control Schemes over Low-Earth-Orbit (LEO) Satellite Networks}},
    year = {2023},
    booktitle = {Proc of AINTEC},
}

@article{BRUHN2023109609,
    title = {Performance and improvements of TCP CUBIC in low-delay cellular networks},
    journal = {Computer Networks},
    year = {2023},
    author = {Philipp Bruhn and Mirja Kühlewind and Maciej Muehleisen},

}

@inproceedings{starrynet,
    author = {Zeqi Lai and Hewu Li and others},
    title = {{StarryNet}: Empowering Researchers to Evaluate Futuristic Integrated Space and Terrestrial Networks},
    booktitle = {NSDI 23},
    year = {2023},
}

@misc{xeoverse,
      title={xeoverse: A Real-time Simulation Platform for Large LEO Satellite Mega-Constellations}, 
      author={Mohamed M. Kassem and Nishanth Sastry},
      year={2024},
      eprint={2406.11366},
      archivePrefix={arXiv},
      primaryClass={cs.NI},
      url={https://arxiv.org/abs/2406.11366}, 
}

@inproceedings{tcp_leo_starlink,
    author = {Garcia, Johan and Sundberg, Simon and Brunstrom, Anna},
    title = {TCP Congestion Control Performance over Starlink},
    year = {2025},
    publisher = {Association for Computing Machinery},
    series = {ANRW '25}
}

@misc{data,
    title = {Code and dataset for the paper titled Taming Congestion in Space: A Study of Congestion Control on LEO Networks},
    url = {https://figshare.com/s/32821d6f5940f72fd946},
}

@inproceedings{leocc,
    author = {Lai, Zeqi and Li, Zonglun and Wu, Qian and Li, Hewu and Li, Jihao and Xie, Xin and Li, Yuanjie and Liu, Jun and Wu, Jianping},
    title = {LeoCC: Making Internet Congestion Control Robust to LEO Satellite Dynamics},
    year = {2025},
    publisher = {Association for Computing Machinery},
    url = {https://doi.org/10.1145/3718958.3750491},
    series = {SIGCOMM '25}
}

@article{hybla,
author = {Caini, Carlo and Firrincieli, Rosario},
title = {TCP Hybla: a TCP enhancement for heterogeneous networks},
year = {2004},
issue_date = {September 2004},
publisher = {John Wiley \& Sons, Inc.},
doi = {10.1002/sat.799}
}

\end{document}